\newcommand{\be}{\begin{eqnarray}}
\newcommand{\ee}{\end{eqnarray}}
\newcommand{\nn}{\nonumber}

\documentclass[12pt]{article}

\usepackage{amsfonts,amsmath,amssymb}
\usepackage{graphicx}
\usepackage{cite}
\usepackage{bm}

\begin{document}

\renewcommand{\thefootnote}{\fnsymbol{footnote}}

\vskip 15mm

\begin{center}

{\Large Crypto-Hermiticity of nonanticommutative theories}

\vskip 4ex

A.V. \textsc{Smilga},

\vskip 3ex

\textit{SUBATECH, Universit\'e de
Nantes,  4 rue Alfred Kastler, BP 20722, Nantes  44307, France
\footnote{On leave of absence from ITEP, Moscow, Russia.}}
\\
\texttt{smilga@subatech.in2p3.fr}
\end{center}

\vskip 5ex

\begin{abstract}
\noindent We note that, though nonanticommutative  deformations of Minkowski supersymmetric
theories do not respect the  reality condition and seem to lead to non-Hermitian Hamiltonians $H$, the
latter belong to the class of crypto-Hermitian (or quasi-Hermitian) Hamiltonians
having attracted recently a considerable attention.
  They can be made manifestly Hermitian via the similarity transformation
$H \to e^R H e^{-R}$ with a properly chosen $R$. The deformed model enjoys {\it the same} supersymmetry
algebra as the undeformed one though it is difficult in some cases to write explicit expressions for a half 
of supercharges. The deformed SQM models make perfect sense. It is not clear whether it is also
the case for NAC Minkowski field theories --- the conventionally defined $S$--matrix is not unitary there.
\end{abstract}

\renewcommand{\thefootnote}{\arabic{footnote}}
\setcounter{footnote}0
\setcounter{page}{1}

\section{Introduction}
Supersymmetric models with nonanticommutative (NAC) deformations \cite{Seiberg} have
recently attracted  a considerable
interest. The main idea is that the odd superspace coordinates $\theta^\alpha$ and
$\bar \theta^{\dot \alpha}$
are not treated as strictly anticommuting anymore, but involve  non-vanishing
anticommutators \cite{Nonc} \footnote{In other words, the original Grassmann algebra of the odd coordinates
is deformed into a {\it Clifford} algebra.}.
In  original Seiberg's paper and in many subsequent works (see e.g. \cite{Euclid,Ito} and
references therein), the deformation
is performed in Euclidean
rather than Minkowski space-time. The reason is that in Minkowski space it seems  impossible
to preserve
both supersymmetry and reality of the action after deformation, still retaining simple properties of
the corresponding $\star$-product (e.g., associativity and nilpotency).
As discussed in Ref.\cite{Seiberg}, Euclidean
NAC theories are of interest in stringy perspectives.
An interesting question
is whether NAC theories are meaningful by themselves, leaving aside the issue
of their relationships
with string theory.
In other words --- whether it is possible
to consistently define them in Minkowski
space. 

We argue that the answer to this question is at least partially positive. Namely, we will show that,
for NAC theories put in finite spatial box,
one can  introduce a Hamiltonian with real spectrum and find
a unitary finite time evolution operator. However, $S$-matrix obtained after projecting this operator onto conventionally
defined $|in\rangle$ and $|out\rangle$ asymptotic states {\it is} not unitary.

Our consideration is  based on the analysis of
two SQM models --- {\it (i)} an interesting 1--dimensional NAC model constructed in a recent paper of Aldrovandi
and Schaposnik \cite{shap} and {\it (ii)} the model obtained from NAC Wess-Zumino model by dimensional reduction.

In Ref.\cite{shap}, NAC deformations of the conventional Witten's supersymmetric quantum mechanics (SQM)
model \cite{Witten} were studied in the chiral basis. In this case,
the deformation operator commutes with the supercharge $Q$, but does not commute with $\bar Q$. However,
Aldrovandi and Schaposnik noticed the presence of the second supercharge $\bar {\cal Q}$ that commutes with
the Hamiltonian. On the other hand, $Q$ and    $\bar {\cal Q}$ seem not to be adjoint to each other
and the deformed Hamiltonian  seems to  lack  Hermiticity.

Our key observation \cite{Ivanov} is that, in spite of having a complex appearance, this Hamiltonian is actually
Hermitian in disguise. One can call it ``crypto-Hermitian'' (or ``cryptoreal''). It belongs to the
class of Hamiltonians having attracted recently a considerable attention (the Hamiltonians of this kind are known since mid-seventies 
\cite{bratmog}, but these studies received great impetus after the beautiful paper \cite{Bender}). One of the simplest examples is
 \be
\label{ix3}
H \ =\ \frac {p^2 + x^2}2 + igx^3\,.
 \ee
In spite of the manifestly complex potential, it is possible to endow the Hamiltonian (\ref{ix3})
with a properly defined Hilbert space such that the spectrum of $H$ is real. The clearest way to see
this is to observe
the existence of the operator $R$ such that the conjugated Hamiltonian
  \be
\label{conjug}
  \tilde H \ = \ e^{R} H e^{-R}
  \ee
 is manifestly
self-adjoint \cite{turok}. The explicit form of $R$ for the Hamiltonian (\ref{ix3}) is
\footnote{Actually, what is written here is the Weyl symbol of the operator $R$. The expression
for a contribution
to the quantum operator corresponding to a monomial $\sim p^n x^n$ in its Weyl symbol is a properly
symmetrized structure, $px \to (1/2)(\hat p x + x \hat p)$, $x^2 p
\to (1/3)( x^2 \hat p + \hat p x^2 + x\hat p x)\,$, etc.}
 \be
\label{Rix3}
  R = g\left( \frac 23 p^3 + x^2 p \right) - g^3 \left( \frac {64}{15} p^5 + \frac {20}3 p^3 x^2 +
4px^4 - 6p \right) + O(g^5)\ .
 \ee
The rotated Hamiltonian is
 \be
 \label{hix3}
\tilde H \ =\ \frac{p^2 + x^2}2 + g^2 \left( 3p^2x^2 + \frac {3x^4}2 - \frac 12 \right) + O(g^4)\ .
 \ee
The (real) spectrum of $\tilde H$ (and $H$) can be found to any order in g in the perturbation theory,
and also non-perturbatively.

 We will see that in  the case of the Aldrovandi-Schaposnik Hamiltonian, there also exists
the operator $R$  making the Hamiltonian Hermitian. The rotated supercharges $e^R Qe^{-R}$ and
$ e^R \bar {\cal Q} e^{-R}$ are Hermitian-conjugated. Such an operator must exist also for the NAC WZ model.

\section{Aldrovandi-Schaposnik model}

The simplest SQM model \cite{Witten} involves a real supervariable
\be
\label{X}
X(\theta, \bar \theta, t) \  = \ x(t) + \theta \psi(t)
+ \bar \psi(t) \bar \theta  + \theta \bar \theta F(t)\ .
 \ee
The action is
 \be
\label{LSQM}
 S \ =\ -\int dt \, d^2\theta \left[ \frac 12 (DX) (\bar D X)  + V(X) \right],
 \ee
with the convention $\int d^2\theta \, \theta \bar\theta = 1\,$. Here $V(X)$ is the superpotential
and $D, \bar D$ are covariant derivatives. Bearing in mind the
deformation coming soon, we will choose their left chiral basis representation
 \be
\label{DDbar}
 D \ =\ \frac \partial{\partial \theta}  - 2i  \bar\theta \frac \partial {\partial t} \, ,\ \ \ \
 \ \ \bar D \ =\
- \frac \partial{\partial \bar \theta} \ .
 \ee
Here $t = \tau -i\theta\bar\theta\,$ and $\tau$ is the real time coordinate of the central basis.
Asymmetry between $D$ and $\bar D$ makes the Lagrangian following from (\ref{LSQM}) complex,
 \be
\label{Lnedef}
 L \ =\ -i\dot x F - \frac{\partial V(x)} {\partial x} F + \frac 12 F^2 + i \bar \psi \dot \psi +
\frac {\partial^2 V(x)}{\partial x^2}
\bar \psi \psi \, ,
 \ee
but one can easily make it real, rewriting it in terms  of $\tilde F = F - i\dot x$
and subtracting a total derivative.
This corresponds to going
over to the central basis from the chiral one.

The deformation is introduced by postulating non-vanishing anticommutators
 \be
\label{deform}
\{\theta, \theta\} = C,\ \ \ \ \{\bar\theta, \bar\theta\} = \bar C, \ \ \ \ \ \{\theta, \bar \theta\}
= \tilde C \, .
  \ee
 The deformed action  involves star products,
   \be
\label{Ldeform}
 S \ =\ -\int dt \, d^2\theta \left[ \frac 12 (D \star X)\star (\bar D \star X)
 +  V_{\star}(X) \right],
 \ee
where
 \be
\label{star}
 X\star Y \ =\  \left. \exp \left\{ - \frac C2 \frac {\partial^2 }{\partial \theta_1 \partial \theta_2}
  - \frac {\bar C}2
\frac {\partial^2}{\partial \bar \theta_1 \partial \bar \theta_2 }
  - \frac {\tilde C}2 \left(
\frac {\partial^2}{\partial  \theta_1 \partial  \bar \theta_2 } +
\frac {\partial^2}{\partial  \bar \theta_1 \partial   \theta_2 } \right) \right\} X(1) Y(2) \right|_{1=2}
 \ee
and $V_{\star}(X)$ is obtained from $V(X) = \sum_n c_n X^n$ by substituting $X^2 \to X_\star^2 \equiv
X\star X,\ X^3 \to X_\star^3 \equiv X\star X\star X$, etc
in its Taylor expansion. The star product  is associative.

The component expression for the deformed Lagrangian is the same as in Eq. (\ref{Lnedef}), with $V(x)$
being substituted by \cite{Alv,shap}
 \be
\label{Vtildint}
 \tilde V(x, F) \ =\ \int_{-1/2}^{1/2} d\xi \, V(x + \xi c F)\ ,
 \ee
where
 \be
\label{c2}
c^2 =  \tilde C^2 - C\bar C
 \ee
 is the relevant deformation parameter.
    If $\bar C$ is conjugate to $C$ and $\tilde C$ is real,
 $c^2$ is also real.
Note, however, that one may, generally speaking, lift the condition that $\theta$ and $\bar \theta$
are conjugate to each other, in which
case $C,\bar C$ and $\tilde C$ can take arbitrary values. We still require the reality of $c^2$.
The crypto-Hermiticity of the deformed Hamiltonian discussed below is fulfilled under this condition.

 In the simplest nontrivial case,
$V(X) = \lambda X^3/3$,
 \be
\label{Vtild}
 \tilde V(x, F) \ =\ \frac {\lambda x^3}3 + \frac {\lambda c^2xF^2}{12} \ .
 \ee
The corresponding canonical Hamiltonian is
 \be
\label{Hamdef}
H \ =\ \frac {p^2}2 + i \frac {\partial \tilde V}{\partial x} p
-  \frac {\partial^2 \tilde V}{\partial x^2} \bar \psi \psi\,,
 \ee
with $p = -iF$. The deformed Lagrangian and Hamiltonian look inherently complex. Obviously, the complexities
now cannot be removed by simply
going from the chiral to the central basis.

In the chiral basis, the supercharges are represented by the following superspace differential operators,
\be
\label{QQbar}
 Q \ =\ \frac \partial{\partial \theta} \, ,\ \ \ \ \ \ \bar Q \ =\
- \frac \partial{\partial \bar \theta}  - 2i  \theta \frac \partial {\partial t}\ .
 \ee
 Note that the star product operator (\ref{star}) still commutes with $Q$ (in  other words, the Leibnitz rule
$Q\star(X\star Y) = (Q\star X)\star Y + X\star (Q\star Y)$ still holds), but not with $\bar Q$.
That means that the deformed
action (\ref{Ldeform})
is still invariant with respect to the supersymmetry transformations generated by $Q$, but not $\bar Q$.
The $Q$-invariance implies the existence of the conserved N\"other supercharge whose component phase space
expression is simply
 \be
\label{Qcomp}
 Q \ =\ \psi p\ .
 \ee
As was observed in \cite{shap}, there is another Grassmann-odd operator commuting with the Hamiltonian.
It reads
 \be
\label{Qbarcomp}
 \bar{\cal Q} \ =\ \bar \psi \left( p + 2i \frac{\partial \tilde V}{\partial x} \right).
 \ee
The standard SUSY algebra
\be
\label{algebra}
 Q^2 = \bar {\cal Q}^2 = 0,\ \ \ \ \ \ \ \{Q, \bar{\cal Q} \} = 2H
 \ee
holds, but,
naively, $\bar{\cal Q}$ is not adjoint to $Q$ and $H$ is not Hermitian.

Let us show now that the Hamiltonian (\ref{Hamdef}) is in fact cryptoreal. Consider for simplicity only the case
(\ref{Vtild}). We have,
 \be
H \ = \ \frac {p^2}2 + i\lambda p x^2 - i\beta p^3 - 2\lambda  x \bar \psi \psi \ ,
 \ee
where $ \beta = \lambda c^2/12$.

It is convenient to treat $\lambda$ and $\beta$ on equal footing and to get rid of
the complexities $\sim ipx^2$ and
$\sim i p^3$ simultaneously. The operator $R$ doing this job is
 \be
\label{RSQM}
R  \ =\ - \frac {\lambda x^3}3 + \beta x p^2 - 2 \lambda \beta x^2 \bar \psi \psi + \ldots \ ,
 \ee
where the dots stand for the terms of the third and higher order in $\lambda$ and/or $\beta$. The conjugated
Hamiltonian is
 \be
\label{hSQM}
\tilde{H} = e^R H e^{-R} \ =\ \frac {p^2}2 - 2\lambda x \bar\psi \psi + \frac 12 [ \lambda^2 x^4 + 3\beta^2 p^4]
+ \frac{1}{2}\lambda\beta + O(\lambda^3,\beta^3, \lambda^2 \beta, \lambda \beta^2)\,.
 \ee
It is Hermitian.
The rotated supercharges are
 \be
\label{QHerm}
 \tilde Q = e^R Qe^{-R} = \psi[p - i(\lambda x^2 - \beta p^2) + \lambda \beta x^2 p - \beta^2 p^3
 + \ldots]\,, \nonumber \\
\tilde {\bar Q} = e^R \bar{\cal Q} e^{-R} = \bar\psi [p + i(\lambda x^2 - \beta p^2)  + \lambda \beta x^2 p +
3\beta^2 p^3 + \ldots ]\,.
  \ee
We observe that they are still not adjoint to each other. To make them mutually adjoint
to the considered order in
$\beta, \lambda\,$, one should add to the operator $R$ one more term
\be
R \; \Rightarrow \; \hat{R} = R - 2\beta^2p^2\bar\psi \psi\,.
\ee
It is easy to see that this modification does not change the rotated Hamiltonian in the considered order,
but ensures the rotated supercharges to be manifestly adjoint to each other
 \be
\label{QHerm1}
 \hat{Q} = e^{\hat{R}} Qe^{-\hat{R}} = \psi[p - i(\lambda x^2 - \beta p^2) + \lambda \beta x^2 p
 + \beta^2 p^3 + \ldots]\,, \nonumber \\
\hat{\bar Q} = e^R \bar{\cal Q} e^{-R} = \bar\psi [p + i(\lambda x^2 - \beta p^2)  + \lambda \beta x^2 p +
\beta^2 p^3 + \ldots ]\,.
  \ee
By construction, the operators $\hat{Q}, \hat{\bar Q}$ and $\tilde{H}$ satisfy the standard
algebra (\ref{algebra}). We see that the requirement of the mutual adjointness of supercharges
is to some extent more fundamental than that of the Hermiticity of the Hamiltonian ---
the latter does not strictly fix the rotation operator $R$ while the former does.

 One can be convinced, order by order in $\beta, \lambda\,$, that complexities in $H$ can
be successfully rotated away also in higher orders (with simultaneously restoring the mutual conjugacy
of the supercharges), and this is also true for higher powers $N > 3$ in $V(X) \sim X^N$ and
hence for any analytic superpotential \footnote{It would be worth being aware of the full analytic
proof of this.}.

\section{NAC Wess-Zumino model}

The first example of an anticommutative deformation of a supersymmetric field theory was considered in
Ref.\cite{Seiberg}. Seiberg took the standard Wess-Zumino model
 \be
\label{WZ}
{\cal L} \ =\ \int d^4\theta \, \bar \Phi \Phi + \left[ \int d^2\theta \left( \frac {m\Phi^2}2
+ \frac {g \Phi^3}3
\right) + {\rm c.c} \right] \nn \\
\equiv \ |\partial_\mu \phi|^2 + i \bar \psi \hat {\partial} \psi  - |F(\phi)|^2 
+ \left[ F'(\phi) \psi^2 + H.c. \right]
 \ee
with $F(\phi) = m\phi + g\phi^2$  and deformed it by introducing the nontrivial anticommutator
  \be
\label{Calbet}
 \{\theta^\alpha, \theta^\beta \} \ =\ C^{\alpha\beta}\ ,
 \ee
$  C^{\alpha\beta} =  C^{\beta\alpha}$, in the assumption that all other (anti)commutators vanish,
  \be
\label{drugiekom}
\{\bar \theta^{\dot \alpha}, \bar \theta^{\dot \beta} \} = \{ \theta^{\alpha}, \bar \theta^{\dot \beta} \} =
[\theta^\alpha, x^L_\mu] = [ \bar \theta^{\dot \alpha}, x^L_\mu] = [x^L_\mu, x^L_\nu] = 0\ .
  \ee
Note that this all was written in the {\it chiral} basis,
$x_\mu^L = x_\mu^{\rm central} + i\theta \sigma_\mu \bar \theta $.
 In Ref.\cite{Seiberg}, the space $x_\mu$ was assumed to be Euclidean. We will work in Minkowski space, however,
 and will not be scared
by the appearance of complexities  at intermediate steps.  The Minkowski space deformation
(\ref{Calbet}), (\ref{drugiekom}) is analogous
to the SQM deformation (\ref{deform}) with $\bar C = \tilde C = 0\,$.

The anticommutator
(\ref{Calbet}) introduces a constant self-dual tensor, which explicitly breaks Lorentz invariance.
However, the deformed Lagrangian expressed in terms of the component fields proves still
to be Lorentz invariant. Indeed, it is easy to find that the
 kinetic term $\int d^4\theta \, \bar \Phi \Phi $ is undeformed and the only extra piece comes from
 \be
 \label{F3}
\Delta {\cal L} =    \frac g 3 
\int d^2\theta \, \Phi*\Phi*\Phi - \frac g 3 \int d^2\theta \, \Phi^3 = 
 -   \frac g 3  \,\det \|C\| F^3\ .
 \ee
 It depends only on
the scalar $~\det\|C\|$ and is obviously Lorentz invariant. Adding the usual terms 
$  F(m \phi + g \phi^2)  +  \bar F(m \bar \phi + \bar g \bar\phi^2) $ \ coming from superpotential 
 and $F\bar F$ from the kinetic term, and
expressing $F$ and $\bar F$ via
$\phi$ and $\bar\phi$, we see that the undeformed potential $|m\phi + g \phi^2|^2$
acquires an extra holomorphic contribution $\propto g(m\bar\phi + \bar g \bar\phi^2)^3$.

When  $g = 0$,  the undeformed model is free and so is deformed one. 
The interacting model is deformed, however, in a nontrivial way \cite{cryptosusy}. Contrary to our original hope \cite{Ivanov},
the spectrum is shifted.
 To see this explicitly, let us consider the dimensionally reduced system and 
assume that the fields do not depend on 
spatial coordinates. The reduced  Hamiltonian is  
  \be
\label{Ham}
H \ =\ \bar \pi \pi +  \bar \phi  \phi +   g \phi^2 \bar \phi +  \bar g \bar \phi^2 \phi   + 
g\bar g \bar \phi^2 \phi^2  -(1+ 2g\phi) \psi_1 \psi_2 - (1 + 2 \bar g \bar \phi) \bar \psi_2 \bar \psi_1 
\nonumber \\
  + \beta ( \bar \phi + \bar g^2 \bar \phi^2)^3 
 \ee
with $\bar \psi_\alpha \equiv \partial/\partial \psi_\alpha$ and $\beta = g \det\|C\|/3$ being the deformation
parameter. For simplicity, we have set $m=1$. 

\begin{figure}[h]
   \begin{center}
 \includegraphics[width=4.0in]{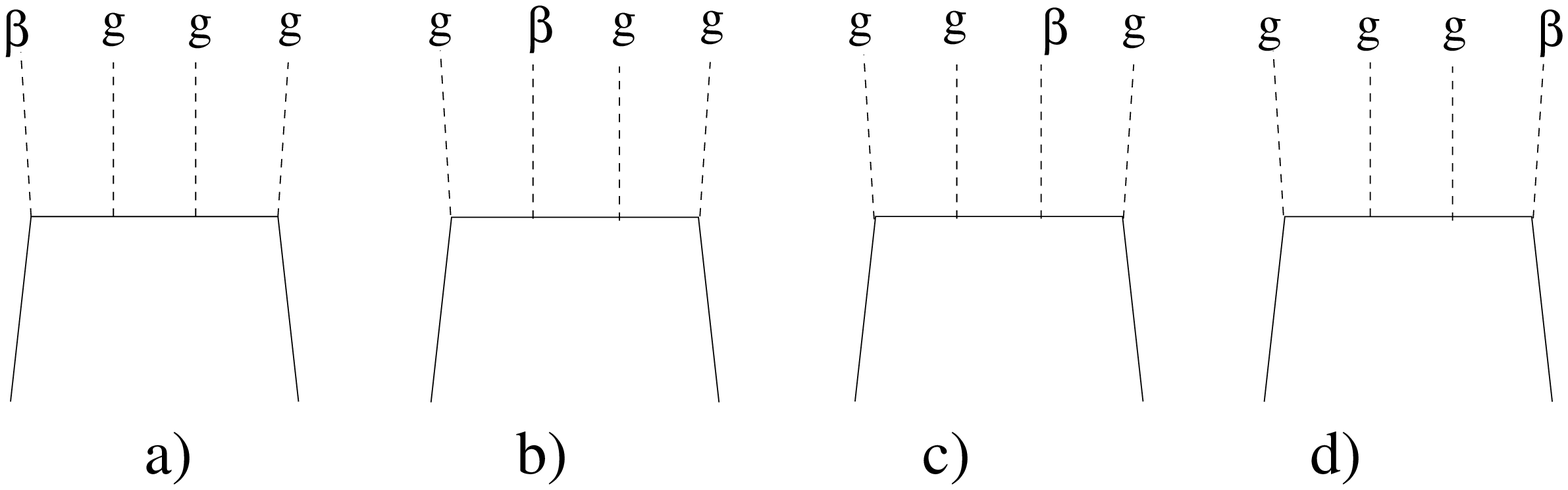}
    \end{center}
\caption{Graphs contributing to the energy shift $\propto \beta  g^3$}
\label{bggg}
\end{figure}

 The wave functions for this Hamiltonian have four components, being  represented as
 \be
\label{Psi}
 \Psi(\phi, \bar\phi, \psi_\alpha) = A(\bar\phi, \phi)  + 
B_\alpha (\bar\phi, \phi)  \psi_\alpha + C (\bar\phi, \phi) \psi_1\psi_2\ .
 \ee
  In the undeformed case, the Hamiltonian (\ref{Ham}) admits conserved supercharges
 \be
\label{superQ}
Q_\alpha = \pi\psi_\alpha + i\epsilon_{\alpha\gamma} \bar \psi_\gamma ( \bar \phi  +  \bar g \bar \phi^2 )\, ,  \nonumber \\
\bar Q_\beta = \bar \pi \bar \psi_\beta - i\epsilon_{\beta\delta} \psi_\delta (\phi +  g \phi^2 )
 \ee
with  $\epsilon_{12} = 1$. They satisfy the usual ${\cal N} = 2$ SQM algebra
 \be
\label{alg}
\{Q_\alpha, Q_\beta \} = \{\bar Q_\alpha, \bar Q_\beta \} = 0, \ \ \ \{Q_\alpha, \bar Q_\beta \} = H\delta_{\alpha\beta}
 \ee
The spectrum of the undeformed Hamiltonian involves a single vacuum state, while the excited
states come in quartets: there is a  quartet of states of energy $1$, two quartets of energy $2$, etc.
One can check by an explicit perturbative calculation (It is a fourth order calculation. The relevant graphs are shown
in Fig. \ref{bggg}) that the ground state of the perturbed Hamiltonian
still has the zero energy, while the energies of the excited states are shifted. For example, for the first excited
quartet, 
 \be
\label{DE1}
 \Delta E_1 = - \frac {155}{36} \beta  g^3  \ + {\rm higher\  order\  terms} \ .
 \ee 
If $\beta  g^3$ is real, the energy shift is also real. 

 It is important that the  spectrum keeps
its structure dictated by the supersymmetry (\ref{superQ}) and consists of degenerate quartets.
This means that the  deformed model still enjoys full supersymmetry of the undeformed theory and 
the algebra (\ref{superQ}) still holds.

Speaking of the supercharge $ Q_\alpha$, it is still given by the expression in 
Eq.(\ref{superQ}) , which commutes
with the  deformed Hamiltonian. On the other hand, the commutator of the undeformed supercharge $\bar Q_\alpha$ with the deformed 
Hamiltonian does not vanish. In contrast to the Aldrovandy-Schaposnik model considered in the previous section,
we cannot write a simple expression for the deformed supercharge. By no means can it be obtained by complex conjugation
of the supercharge  $ Q_\alpha$. Indeed, a pair of complex conjugate supercharges would mean Hermiticity of Hamiltonian, 
but the Hamiltonian (\ref{Ham}) is not manifestly Hermitian. 
The fact that its spectrum is real (when $\beta, g$ are real) tells, however, that the Hamiltonian is crypto-Hermitian
in the same sense as the Aldrovandy-Schaposnik Hamiltonian is. In particular, the operator $R$ rotating the Hamiltonian
to the manifestly Hermitian form should exist.

Even though explicit expressions for $\bar Q_\alpha$ are not known, one can argue 
that the quartet supersymmetric structure of the spectrum must hold without making explicit calculations. It 
can be reconstructed 
(at least, perturbatively \footnote{It would be very interesting to study the spectrum of the deformed 
Hamiltonian numerically. One cannot  
exclude a possibility that exceptional points \cite{Heiss} in the  space of couplings   appear such that 
the supersymmetric structure of the spectrum would be lost for large enough values of $\beta, g$.}) 
using only $ Q_\alpha$ and not $\bar Q_\alpha$. 
 Indeed, for each supersymmetric quartet of the eigenstates of the free Hamiltonian $H_0$,  
a member $\Psi$ annihilated by the action of $\bar Q_\alpha$, but not $ Q_\alpha$, can be chosen. 
 Three other members of the quartet are $ Q_{1,2}\Psi$ and
$ Q_1  Q_2 \Psi$. Let $\tilde \Psi$ be the corresponding eigenstate of the full Hamiltonian 
(when $\beta$ and $g$ are small, one can be sure
that such state exists). Then $\tilde \Psi$,  $ Q_\alpha \tilde \Psi$, and $ Q^2 \tilde\Psi$ 
represent a quartet of degenerate eigenstates of the interacting
deformed Hamiltonian. Once the states are known, the matrix elements of $\bar Q_\alpha$ can be defined 
to be equal to the corresponding matrix elements
in the free undeformed basis multiplied by $\sqrt{E_n^{\rm exact}/E_n^{\rm free}}$. 

What conclusions concerning NAC field theories can be made on the basis of this analysis ?
If we put the theory in a finite spatial box and be interested in the spectrum of the Hamiltonian
thus obtained, its properties should be similar to the properties of the dimensionally reduced Hamiltonian:
 \begin{itemize}
\item The ground state energy(ies) is(are) still zero (if supersymmetry is not spontaneously broken)
and the $2^{\cal N}$ degeneracy  of the excited spectrum states should be kept.
 \item For certain values of the deformation parameters and the couplings, the spectrum of the deformed Hamiltonian
 should enjoy crypto-Hermiticity property.
\end{itemize}

 However, the main question we usually ask in field theories is not what are the spectra of their 
finite box Hamiltonians, but what are their $S$-matrices --- the matrix elements of the evolution operator
between the asymptotic $|in\rangle$ and $|out\rangle$ states. For NAC theories, the complexity of Lagrangian
strikes back at this point: for {\it conventionally} defined asymptotic states, the $S$-matrix for, say
NAC Wess-Zumino model is not unitary (see Ref.\cite{cryptosusy} for more detailed discussion).  

This means that NAC theories obtained by deformation of {\it interacting} SUSY theories cannot be attributed
a {\it conventional} physical meaning. More studies of this question are necessary. Maybe even if $S$-matrix of the
theory is not unitary, unitarity of its finite time finite box evolution operator (that follows from crypto-Hermiticity
of the Hamiltonian) suffices to  make the theory meaningful ? A positive answer to this question would mean a breakthrough
in understanding not only  NAC theories, but also theories with higher derivatives in the Lagrangian. In Ref.\cite{TOE},
we argued that the fundamental Theory of Everything may be a theory of this kind. We address the reader to this paper and 
also to the papers \cite{brmog1} for discussions and speculations on this subject.

\section*{Acknowledgements}
It is a pleasure to thank Evgeny Ivanov for collaboration (this talk is based in a considerable extent on
our paper \cite{Ivanov}) and numerous fruitful discussions.

\end{document}